\documentstyle{article}
\newcommand{\vect}[1]{\mbox{\bf #1}}
\newcommand{\vecs}[1]{\mbox{\scriptsize \bf #1}}

\newcommand{\lw}[1]{\smash{\lower 1.5ex\hbox{#1}}}
\begin{document}
%
%
\title{{\Large \bf Improved Coulomb Correction Formulae}
\vspace{-4mm}\\
  {\Large \bf for Bose-Einstein Correlations }}
  \author{M.~Biyajima$^1$\thanks{e-mail: minoru44@jpnyitp.bitnet},
T.~Mizoguchi$^2$, T.~Osada$^3$ and G.~Wilk$^4$\thanks{e-mail:
wilk@fuw.edu.pl}\\
  {\small $^1$Department of Physics, Faculty of Liberal Arts,
Shinshu University, Matsumoto 390, Japan}\vspace{-2mm}\\
  {\small $^2$Toba National College of Maritime Technology,
Toba 517, Japan}\vspace{-2mm}\\
  {\small $^3$Department of Physics, Tohoku University,
Sendai 980, Japan}\vspace{-2mm}\\
  {\small $^4$Soltan Institute for Nuclear Studies, Zd-PVIII,
Ho\.za 69, PL-00-681 Warsaw, Poland}}
\date{}
\maketitle
%
%
\begin{abstract}
We present improved Coulomb correction formulae for Bose-Einstein
correlations including also exchange term and use them to calculate
appropriate correction factors for several source functions. It is
found that Coulomb correction to the exchange function in the
Bose-Einstein correlations cannot be neglected.
\end{abstract}
{\it SULDP-1995-3~~~~~TU477~~~~~SINS-1995-1}
\vspace{1cm}
{\bf Introduction:} Recently many experimental groups have
investigated Bose-Einstein correlations (BEC) in high energy hadronic
collisions \cite{agababyan}, $e^+e^-$ annihilations \cite{DELPHI} and
heavy-ion collisions~\cite{boggild,na4494}. The high quality of data
obtained already (and expected in near future, especially for
heavy-ion collisions) make the analysis of experimental results
sensitive to all possible corrections, especially to those due to
final state interactions because of the Coulomb interactions and the
strong  interactions \cite{biya95}. In the present letter we shall
concentrate  on the Coulomb corrections only. \\

Several authors have calculated Coulomb corrections to the BEC by their
own methods~\cite{gkw,pratt86,gersch,boal,bowler91} but this problem is
still controversial \cite{bowler91,anchishkin94}. In this letter we
shall concentrate on the formula provided by Bowler~\cite{bowler91},
because it includes all orders of the parameter $\eta $ (defined
below). He demonstrated that widely used zero range Gamow factor
substantially overestimates Coulomb corrections (and hence also the
true magnitude of BEC). However, Bowler did not calculate the Coulomb
correction to the exchange function present in BEC formulae. We
investigate therefore this problem and in particular: (1) we calculate
the Coulomb corrections using the exact formula (to all orders in
parameter $ \eta $) and (2) we use its approximate form (retaining only
terms linear in $ \eta $) to calculate explicit Coulomb wave function
in this approximation \cite{pratt86,biya94,biya94d} and with its help
we provide formulae for Coulomb corrections for the exchange function
in the BEC for several source functions and compare some of them with
those presented in ref.\cite{bowler91}. We demonstrate that the
presence of exchange term diminishes correction factor even more than
anticipated in \cite{bowler91} and therefore it should be included in
analysis of experimental data.
\vspace{5mm}\\
{\bf Theoretical calculation of BEC with Coulomb wave function:}
To write down an amplitude $A_{12}$ satisfying Bose-Einstein statistics
it is convenient to decompose the wave function of identical (charged
in our case) bosons with momenta $p_1$ and $p_2$ into the wave function
of the center-of-mass system (c.m.) with total momentum $P =
\frac{1}{2}(p_1 + p_2)$ and the inner wave function with relative
momentum $Q = (p_1 - p_2) = 2q$. It allows us to express $A_{12}$ in
terms of the confluent hypergeometric function $\Phi$ \cite{schiff}:
\begin{eqnarray}
 A_{12} &=& \frac 1{\sqrt{2}} [ \Psi(\vect{q},\vect{r}) +
\Psi_S(\vect{q},\vect{r}) ]\:,\\
 \Psi(\vect{q},\vect{r}) &=& \Gamma(1+i\eta)e^{-\pi \eta/2}
e^{i\vecs{q}\cdot\vecs{r}}
\Phi(-i\eta;1;iqr(1 - \cos \theta))\:,\nonumber\\
 \Psi_S(\vect{q},\vect{r}) &=& \Gamma(1+i\eta)e^{-\pi \eta/2}
e^{-i\vecs{q}\cdot\vecs{r}}
\Phi(-i\eta;1;iqr(1 + \cos \theta))\:,\nonumber
\end{eqnarray}
where $r = x_1 - x_2$ and the parameter $\eta = m\alpha/2q$.  Using
now the Kummer's first formula for the confluent hypergeometric
functions that appears in the cross term of $\mid A_{12}\mid$,
$$\Phi(\alpha;\gamma;z)=e^{z} \Phi(\gamma - \alpha;\gamma;-z),$$
we calculate first the {\it exact} formula of Coulomb correction (i.e.,
the one that is exact to all orders of parameter $\eta$) {\it including
also the exchange function} in BEC. Namely, assuming factorization in
the source functions, $\rho(r_1)\rho(r_2) = \rho(R)\rho(r)$ (here $R =
\frac{1}{2}(x_1 + x_2)$), one obtains the following expression for
theoretical BEC formula:
\begin{eqnarray}
N^{\pm \pm}/ N^{BG} &=& \frac 1{G(q)} \int \rho(R)
  d^3R \int \rho(r) d^3r |A_{12}|^2 \nonumber\\
 &=& \sum_{n=0}^{\infty} \sum_{m=0}^{\infty} \frac{(-i)^n(i)^m}{n+m+1}
(2q)^{n+m}I_R(n,m) 
A_n A_m^* \nonumber\\
&\times& \left[ 1+ \frac{n!m!}{(n+m)!}
\left( 1+\frac{n}{i\eta}\right)
\left( 1-\frac{m}{i\eta}\right) \right ] ,\label{eq:result}
\end{eqnarray}
\noindent
where $G(q) = 2\pi \eta /(e^{2\pi \eta} - 1)$ is the Gamow factor and\\
$$ I_R(n,m) = 4 \pi \int dr\, r^{2 + n + m} \rho (r),\qquad
A_n = \frac{\Gamma(i\eta + n)}{\Gamma(i\eta)}\frac{1}{(n!)^2}.
$$\\
The second term in the squared parenthesis in eq.(\ref{eq:result}) is
a new term which is due to the exchange function in the BEC. The above
expression represents our main result. However, it has disadvantage
that it is difficult to separate in it the exchange function and its
Coulomb correction. Therefore, in order to know the separate
contributions we must either subtract from it the exchange function or
to use the {\it approximate} formula for the Coulomb correction. To
calculate it we expand function $\Phi$ in powers of $\eta$ and
retaining only linear terms we get
\begin{eqnarray}
  &&\Phi(-i\eta;1;ix) = 1 + \eta \mbox{Si}(x) -i\eta
(\mbox{Ci}(x) -
\gamma_{\mbox{\tiny E}} - \ln(x)) + O(\eta^2)\:,\nonumber \\
  &&\Phi(-i\eta;1;i\tilde{x}) =  1 + \eta \mbox{Si}(\tilde{x})
-i\eta (\mbox{Ci}(\tilde{x})
- \gamma_{\mbox{\tiny E}} - \ln(\tilde{x}))  + O(\eta^2)\: \nonumber
\end{eqnarray}
where $x = iqr(1 - \cos \theta)$, $ \tilde{x}=iqr(1 - \cos \theta)$,
$\gamma_{\mbox{\tiny E}}$ is the Euler's number and $\mbox{Si}(x)$ and
$\mbox{Ci}(x) - \gamma_{\mbox{\tiny E}} - \ln(x)$ are the sine and
cosine integral, respectively, defined as:
$$\mbox{Si}(x) = \sum_{n=0}^{\infty} \frac{(-1)^n x^{2n + 1}}
{(2n +1)!(2n +1)}\:,\qquad
 \mbox{Ci}(x) - \gamma_{\mbox{\tiny E}} - \ln(x) = \sum_{n=1}^
{\infty}
\frac{(-1)^n x^{2n}}{(2n)!(2n)}\:.$$
\noindent
With such approximated $\Phi$'s we obtain the following {\it
approximate} formula for BEC with Coulomb corrections included:
\begin{eqnarray}
  N^{(\pm\pm)}/N^{BG} &=& \frac 1{G(q)} \int \rho(R)
                 d^3R \int \rho(r) d^3r |A_{12}|^2 \nonumber\\
                      &=& { I_1(q)  +  I_2(q) } \label{eq:Ours1}\\
  &=& 1 + \delta_{1C} + \delta_{EC} + E_{2B}\nonumber\\
  &=& (1 + \delta_{1C} + \delta_{EC})\left( 1 +
\frac{E_{2B}}{1 + \delta_{1C} + \delta_{EC}}
\right)\:\label{eq:Ours}
\end{eqnarray}
where ($A = 2qr$)
\begin{eqnarray}
  I_1(q) &=& 4\pi \int \rho(r) r^2 dr \left\{ 1 + \eta
\sum_{n=0}^{\infty} \frac{(-1)^n (qr)^{2n + 1}}{(2n + 1)!(2n + 1)}
\int_{-1}^{1} (1 - \cos \theta)^{2n+1} d\cos\theta \right\}
\nonumber\\
   &=& 1 + 4\pi\cdot 2\eta\int \rho(r) r^2 dr\sum_{n=0}^{\infty}
\frac{ (-1)^n A^{2n + 1} }{ (2n + 1)!(2n + 1)
(2n + 2) }\nonumber\\
         &=& 1 + \delta_{1C}\:,\\
  I_2(q) &=& 4\pi \int \rho(r) r^2 dr\left\{\frac
{\sin A}{A} +
\eta[\mbox{Sp}(qr) + \mbox{Cp}(qr)]\right\}\nonumber\\
         &=& E_{2B} + \delta_{EC}\:,
\end{eqnarray}
and
\begin{eqnarray}
  &&\mbox{Sp}(qr) = \sum_{n=0}^{\infty}
  \frac {(-1)^{2n+1}(qr)^{2n+1}\Theta(2n+1)}{(2n+1)!(2n+1)}\:
,\nonumber\\
  &&\mbox{Cp}(qr) = \sum_{n=1}^{\infty}
  \frac {(-1)^{2n}(qr)^{2n}\Theta(2n)}{(2n)!(2n)}\:,
\label{eq:SpCp}
\end{eqnarray}
with
\begin{eqnarray}
  &&\Theta(2n+1) = \int (1 - \cos \theta)^{2n+1}
  \cos(A\cos \theta)d\cos \theta\:,\nonumber\\
  &&\Theta(2n) = \mp \int (1 \pm \cos \theta)^{2n}
  \sin(A\cos \theta)d\cos \theta\:. \label{eq:Theta}
\end{eqnarray}
(Notice that $\Theta(1) = \frac {2\sin A}{A},\quad \Theta(3) = \frac
{2^3\sin A}{A} + \frac {12 \cos A}{ A^2 } - \frac {12 \sin A}{ A^3
},\dots $ and $\Theta(2) = \frac {2^2\cos A}{A} - \frac {2^2\sin
A}{A^2}, \dots $). The normalization-like factor $( 1 + \delta_{1C} +
\delta_{EC})$ in eq.(\ref{eq:Ours}) differs from that in
ref.~\cite{bowler91} only by the additional exchange term
$\delta_{EC}$. There is therefore the following (approximate)
relation between our correction factor, $C^{Ours}$, and that of
Bowler~\cite{bowler91}, $C^{Bowler}$:
\begin{equation}
  C^{Ours} = C^{Bowler} + \delta_{EC}\cdot G(q) \label{eq:factor}
\end{equation}
where $G(q)$ is Gamow factor.
\vspace{0.5cm}\\
{\bf Coulomb corrections and source functions:} In order to provide
numerical estimations of our new correction factor we calculate
analytical expressions for necessary ingredients of both exact result
as given by eq.(\ref{eq:result}), $I_R$, and the approximate one
provided by eqs.(\ref{eq:Ours}) and (\ref{eq:factor}),
$\delta^E_{1C},~E_{2B}$  and $\delta_{EC}$ for several different
source functions (superscripts  below denote the type of source
considered):\\
{\bf (S1) Exponential source function, $\rho(r) =
\frac{\beta^3}{8\pi}\exp(-\beta r)$:}\\
Taking into account only leading terms in $\Theta(2n+1)$ and
$\Theta(2n)$ in eqs. (\ref{eq:SpCp}) one has:
\begin{eqnarray}
I_R^E (n,m) &=& \left( \frac{1}{\beta}\right)^{n+m}
\frac{(n+m+2)\,!}{2}, \\
  \delta_{1C}^{E} &=& \eta \sum_{n=0}^{\infty}
\frac{(-1)^n(2n+3)}{2n+1}(2q/\beta)^{2n+1}\:,\\
       E_{2B}^{E} &=& \frac 1{(1+(2q/\beta)^2)^2}\:,\\
  \delta_{EC}^{E} &=& \frac{\eta}2\sum_{n=1}^{\infty}
\frac{(-1)^n(2n+1)}{2n}\frac{(2q/\beta)^{2n-1}}
{(1 + (2q/\beta)^2)^{n+1}}
\cos( (2n+2)\arctan(2q/\beta) )\nonumber\\
     &+& \frac{\eta}2 \sum_{n=0}^{\infty}\frac{(-1)^n(2n+2)}
{2n+1} \frac{(2q/\beta)^{2n}}{(1 + (2q/\beta)^2)^{n+3/2}}
\sin( (2n+3)\arctan(2q/\beta) )\:.\hspace{10mm}
\end{eqnarray}
Our results for this source function are given in Fig. 1. Notice that
there is significant systematic difference between $C^{Bowler}$ and
$C^{Ours}$ correction terms whereas $C^{Ours}$ is practically the same
whenever calculated using exact formula (\ref{eq:result}) or
approximate one (\ref{eq:Ours}) (in the range of $q$ considered the
differences are of the order of $1\%$). \\
{\bf (S2) Gaussian source distribution, $\rho(r) =
\frac{\beta^3}{\sqrt{\pi^3}}\exp(-\beta^2 r^2)$:}\\
For this type of source function we have:
\begin{eqnarray}
I_R^G (n,m) &=& \frac {2}{\sqrt{\pi}}\left(\frac{1}
{\beta}\right)^{n+m}
(n+m+1)~\Gamma\left(\frac{n+m+1}{2} \right),\\
 \delta_{1C}^{G} &=& \frac{4\eta}{\sqrt{\pi}}
\sum_{n=0}^{\infty} \frac{(-1)^n(n+1)!}{(2n+2)!(2n+1)}\left(
\frac{2q}{\beta}\right)^{2n+1}\:,\\
  E_{2B}^{G} &=& \exp\left(-\frac{q^2}{\beta^2}\right)\:,\\
  \delta_{EC}^{G} &=& \frac{2\eta}{\sqrt{\pi}}
\sum_{n=1}^{\infty} \frac{(-1)^nn!}{(2n)!2n}
\left(\frac{2q}{\beta}\right)^{2n-1}
\Phi\left(n+1,\frac 12,-\frac{q^2}{\beta^2}\right)\nonumber\\
  &+& \frac{2\eta}{\sqrt{\pi}}\exp
\left(-\frac{q^2}{\beta^2}\right) \sum_{n=0}^{\infty}
\frac{(-1)^n(n+1)!)}{(2n+1)!(2n+1)}\left(\frac{2q}
{\beta}\right)^{2n+1}
\Phi\left(n+\frac 12,\frac 32,-\frac{q^2}{\beta^2}\right).\hspace{8mm}
\end{eqnarray}
In Fig. 2 we show typical examples of our correction term for
Gaussian source distribution. Also here exact and approximate formulas
lead practically to the same results at lower values of $q$ (compatible
with those in Fig. 1) but differ substantially for larger $q$'s. The
difference between our results and the result obtained using Bowler's
formula (cf. eq.(\ref{eq:Bowler}) below) exists also here.\\
{\bf (S3) Modified Bessel source functions, $\rho(r) =
\frac{\beta^3}{2\pi^2}K_0(\beta r)$ and $\rho(r) =
\frac{\beta^4}{6\pi^2}rK_1(\beta r)$:}\\
In the case of source functions described by the Modified Bessel
functions $K_0$ or $K_1$ \cite{shimoda,mizo94} we have the following
expressions:
\begin{eqnarray}
I_R^{K_0} (n,m) &=& \frac{2^{n+m+2}}{\pi} \left(\frac{1}
{\beta}\right)^{n+m}
\Gamma \left( \frac{n+m+3}{2} \right)\Gamma
\left( \frac{n+m+3}{2} \right),\\
 \delta_{1C}^{K_0} &=& \frac{8\eta}{\pi}
\sum_{n=0}^{\infty} \frac{(-1)^n\{(n+1)!\}^2}{(2n+2)!(2n+1)}
\left(\frac{4q}{\beta}\right)^{2n+1}\:,\\
  E_{2B}^{K_0} &=& \frac 1{(1 + (2q/\beta)^2)^{3/2}}\:,\\
  \delta_{EC}^{K_0} &=& \frac{4\eta}{\sqrt{\pi}}
\sum_{n=1}^{\infty} \frac{(-1)^n}{(2n)!2n}\left(\frac{4q}
{\beta}\right)^{2n-1}
\sum_{m=0}^{\infty} \frac{ (-1)^m\{(n+m)!\}^2 }{\Gamma(m +
\frac 12)m!}\left(\frac{2q}{\beta}\right)^{2m}\nonumber\\
  &+& \frac{2\eta}{\sqrt{\pi}} \sum_{n=0}^{\infty}
\frac{(-1)^n}{(2n+1)!(2n+1)}\left(\frac{4q}{\beta}\right)^{2n+1}
\sum_{m=0}^{\infty} \frac{ (-1)^m\{(n+m+1)!\}^2 }{\Gamma(m +
\frac 32)m!}\left(\frac{2q}{\beta}\right)^{2m}.\hspace{10mm}
\end{eqnarray}
and
\begin{eqnarray}
I_R^{K_1} (n,m) &=& \frac{2^{n+m+3}}{3\pi} \left(\frac{1}
{\beta}\right)^{n+m}
\Gamma \left( \frac{n+m+3}{2}\right)\Gamma \left
( \frac{n+m+5}{2} \right),\\
 \delta_{1C}^{K_1} &=& \frac{16\eta}{3\pi}
\sum_{n=0}^{\infty} \frac{(-1)^n\{(n+1)!\}^2(n+2)}
{(2n+2)!(2n+1)}\left(\frac{4q}{\beta}\right)^{2n+1}\:,\\
  E_{2B}^{K_1} &=& \frac 1{(1 + (2q/\beta)^2)^{5/2}}\:,\\
  \delta_{EC}^{K_1} &=& \frac{8\eta}{3\sqrt{\pi}}
  \sum_{n=1}^{\infty} \frac{(-1)^n}{(2n)!2n}\left(
  \frac{4q}{\beta}\right)^{2n-1}
\sum_{m=0}^{\infty} \frac{ (-1)^m\{(n+m)!\}^2(n+m+1) }
{\Gamma(m +\frac 12)m!}\left(\frac{2q}{\beta}\right)^{2m}
\nonumber\\
  &+& \frac{4\eta}{3\sqrt{\pi}}\sum_{n=0}^{\infty}
\frac{(-1)^n}{(2n+1)!(2n+1)}\left(\frac{4q}{\beta}\right)^
{2n+1}\nonumber\\
&& \times \sum_{m=0}^{\infty}
\frac{ (-1)^m\{(n+m+1)!\}^2(n+m+2) }{\Gamma(m + \frac 32)m!}
\left(\frac{2q}{\beta}\right)^{2m}.
\end{eqnarray}
The correction factors for these source functions are presented
in Fig.~3 which closely resembles Fig. 1a (again, approximate formula
practically does not differ from the exact one whereas both differ
substantially from that calculated according to \cite{bowler91}).
\vspace{0.5cm}\\
{\bf Concluding remarks:} We have calculated the exact and approximate
formulae for Coulomb corrections used in the BEC (including exchange
term) and specified them to the Exponential, Gaussian and Modified
Bessel source functions. The difference between exact and approximate
formulas (eq. (\ref{eq:result}) and eq.(\ref{eq:Ours}), respectively)
turns out to be smaller than 1 \% (at least for $q \leq 0.2$ GeV/c).
However, due to the presence of the new correction for the exchange
term ($\delta_{EC}$ above) our results are systematicall lower than those
derived in \cite{bowler91}. To visualize it better we check if relation
\begin{equation}
   \frac {E_{2B} \delta_{1C}} { \delta_{EC}} = 1 \label{eq:Ratio}
\end{equation}
holds; $E_{2B}$ and $\delta_{1C}$ above are quantities used in ref.
\cite{bowler91}:
\begin{equation}
  N^{\pm\pm}/N^{BG} = ( 1 + \delta_{1C})(1 +  E_{2B}). \label{eq:Bowler}
\end{equation}
It is clear from Fig. 4 that (\ref{eq:Ratio}) does not hold (here for the
standard value of $\beta = 0.2$ GeV used also in \cite{bowler91} but we
have checked that the same is true for the whole possible range of this
parameter). We conclude that correction term $\delta_{EC}$ in
eqs.(\ref{eq:result}) and (\ref{eq:Ours}) cannot be neglected
and that in calculations of Coulomb corrections eq.(\ref{eq:Bowler})
should be replaced by eq.(\ref{eq:Ours}) (or eq.(\ref{eq:result})).\\

In ref.\cite{bowler91} the relative changes of the normalised Fourier
transforms of a given source function, $\frac {\Delta
\tilde{\rho}}{\tilde{\rho}}$, were also introduced and investigated as
yet another estimate of the importance of the Coulomb corrections:
\begin{equation}
  \frac {\Delta \tilde{\rho}}{\tilde{\rho}} = \frac{
    (1 +\delta_{1C})(1+E_{2B})
   - 1}{(1 + E _{2B}) - 1} - 1, \label{eq:E1}
\end{equation}
for the ideal BEC and
\begin{equation}
  \frac {\Delta \tilde{\rho}}{\tilde{\rho}}^R =
   \frac{ x (1 +\delta_{1C})
     (1+E_{2B})   +G^{-1} (1 - x) - 1 }{x[(1 + E _{2B}) - 1]} - 1
     \label{eq:E2}
\end {equation}
for BEC containing contributions of the long lived resonances $\{L\}
=\{\eta, \omega, \eta^{'};c,b\}$ with $1-x$ denoting the fraction of
pairs involving a daughter of $\{L\}$. In Figs. 5 and 6, using the
exact formulae containing also the correction term $\delta_{EC}$, we
compare our results for $\frac {\Delta \tilde{\rho}}{\tilde{\rho}}$ and
$\frac {\Delta \tilde{\rho}}{\tilde{\rho}}^R$  calculated for
Exponential and Gaussian sources for different $\beta$'s (in Fig. 5)
and different $x$'s (in Fig. 6). Contrary to the previous cases,
for these quantities the introduction of Coulomb correction to the
exchange term in BEC practically does not change the results obtained
by using eq.(\ref{eq:Bowler}) from \cite{bowler91}, due to the large
denominator.
\vspace{0.5cm}\\
{\bf Acknowledgements: } The authors would like to thank S. Esumi, S.
Nishimura, and S. D. Pandey for useful correspondences. This work is
partially supported by Japanese Grant-in-Aid for Scientific Research
from the Ministry of Education, Science and Culture (\#. 06640383).
%
%

%
%
\section*{{\large\bf Figure Captions}}
\begin{description}
  \item[Fig.~1. ] Comparison of $C^{Bowler}$ with $C^{Ours}$ for
                  exponential source function and for different
                  choices of parameter $\beta $:
                  (a) $\beta = 0.2 $ GeV. (b) $\beta = 0.1 $ GeV.
  \item[Fig.~2. ] The same as in Fig. 1 but for Gaussian source
                  distribution.
  \item[Fig.~3. ] The same as in Fig. 1 but for Modified
                  Bessel source functions ($K_0$ for (a) and $K_1$
                  for (b)) - only for $\beta = 0.2$ GeV in both cases.
  \item[Fig.~4. ] Examinations of eq. (\ref{eq:Ratio}) for
                  different choices of source functions.
  \item[Fig.~5. ] $ {\Delta \tilde{\rho}}/{\tilde{\rho}}$ (cf.
                  eq.(\ref{eq:E1})) for different $\beta$'s and for
                  Exponential and Gaussian source distributions.
                  In our calculations the exact formulae are used.
  \item[Fig.~6. ] $ {\Delta \tilde{\rho}^R}/{\tilde{\rho}}$
                  (cf. eq.(\ref{eq:E2})) for
                  Exponential and Gaussian source functions and for
                  different values of the fraction of long lived
                  resonances parameter $x$. In our calculations
                  the exact formulae are used.
\end{description}

\begin{thebibliography}{99}
  \bibitem{agababyan}M.~Agababyan et~al. (EHS/NA22 Collaboration),
          {\sl Z.~Phys.} {\bf C59} (1993) 195.
  \bibitem{DELPHI} P.~Abreu et al. (DELPHI Collaboration),
          {\sl Phys.~Lett.} {\bf 323} (1994) 242.
  \bibitem{boggild} H.~B\o ggild et al. (NA44 Collaboration),
          {\sl Phys.~Lett.} {\bf B302} (1993) 510.
  \bibitem{na4494} H. Beker et al. (NA44 Collaboration),
          {\sl Z. Phys.} {\bf C64} (1994) 209.
  \bibitem{biya95} M. Biyajima, T. Mizoguchi, and G. Wilk,
          {\sl Z. Phys.} {\bf C65} (1995) 511.
  \bibitem{gkw} M.~Gyulassy, S~.K.~Kaufmann and L.~W.~Wilson,
          {\sl Phys.~Rev.} {\bf C20} (1979) 2267.
  \bibitem{pratt86} S. Pratt, {\sl Phys.~Rev.} {\bf D33} (1986) 72.
  \bibitem{gersch} H.-U. Gersch, {\sl Z. Phys.} {\bf 327} (1987) 115.
  \bibitem{boal}D.~H.~Boal, C.~-K.~Gelbke and B.~K.~Jennings,
          {\sl Rev.~Mod.~Phys.} {\bf 62} (1990) 553.
  \bibitem{bowler91}M. G. Bowler, {\sl Phys.~Lett.} {\bf B270} (1991)
          69.
  \bibitem{anchishkin94} D. Anchishkin and G. Zinojev, {\it Two-Pion
           Correlation Behaviour in Small Relative Momentum Region},
           BI-TP 94/19 (1994).
  \bibitem{biya94} M. Biyajima and T. Mizoguchi, talk given at the
           workshop on "Evolution from quarks to hadrons" at Yukawa
           Institute for Theoretical Physics. (Oct. 1994).
  \bibitem{biya94d} M. Biyajima and T. Mizoguchi, preprint SULDP-94-9
           (Dec.,1994). Several improper expressions in
           ref.\cite{pratt86} are pointed out and corrected therein.
  \bibitem{schiff} L. I. Schiff, {\it Quantum Mechanics}, 2nd Ed.
           (McGraw-Hill, New York, 1955) p.117.
  \bibitem{shimoda}R.~Shimoda, M.~Biyajima, and N.~Suzuki,
           {\sl Prog.~Theor.~Phys.} {\bf 89} (1993) 697.
  \bibitem{mizo94} T. Mizoguchi, M. Biyajima and T. Kageya,
           {\sl Prog.~Theor.~Phys.} {\bf 91} (1994) 905.
%
\end{thebibliography}
\end{document}